\documentclass[runningheads]{llncs}
\usepackage{graphicx}
\newcommand{\comment}[1]{}
\usepackage{hyperref}
\usepackage[linesnumbered,ruled,vlined]{algorithm2e} 

\usepackage{array}
\begin{document}
\title{Region Specific Optimization (RSO)-based Deep Interactive Registration}

\titlerunning{RSO-based Deep Interactive Registration}
\author{Ti Bai\inst{1}\orcidID{0000-0002-6697-7434} \and
Muhan Lin\inst{1} \and
Xiao Liang\inst{1} \and
Biling Wang\inst{1, 2} \and
Michael Dohopolski\inst{1} \and
Bin Cai\inst{1} \and
Dan Nguyen\inst{1} \and
Steve Jiang\inst{1}\thanks{Corresponding Author: Steve.Jiang@UTSouthwestern.edu}
}
\authorrunning{T. Bai et al.}
%
\institute{Medical Artificial Intelligence and Automation (MAIA) Lab, Department of Radiation Oncology,
UT Southwestern Medical Center, Dallas, TX, USA \and
Department of Statistical Science, Southern Methodist University, Dallas, TX, USA
}

\maketitle              

\begin{abstract}
Medical image registration is a fundamental and vital task which will affect the efficacy of many downstream clinical tasks. Deep learning (DL)-based deformable image registration (DIR) methods have been investigated, showing state-of-the-art performance. A test time optimization (TTO) technique was proposed to further improve the DL models' performance. Despite the substantial accuracy improvement with this TTO technique, there still remained some regions that exhibited large registration errors even after many TTO iterations. To mitigate this challenge, we firstly identified the reason why the TTO technique was slow, or even failed, to improve those regions' registration results. We then proposed a two-levels TTO technique, i.e., image-specific optimization (ISO) and region-specific optimization (RSO), where the region can be interactively indicated by the clinician during the registration result reviewing process. For both  efficiency and accuracy, we further envisioned a three-step DL-based image registration workflow. Experimental results showed that our proposed method outperformed the conventional method qualitatively and quantitatively.

\end{abstract}
\section{Introduction}
The medical image registration task is to find a transformation mapping such that the transformed moving image can be anatomically spatially aligned with the fixed image. Its accuracy is of critical importance since it will affect the effectiveness of many downstream clinical tasks such as organs/tumors' contour propagation in radiotherapy.

Benefiting from the powerful expressiveness of the modern deep learning (DL) models which have witnessed unprecedented success in many fields, the DL methods have been introduced into the medical image registration field, leading to the state-of-the-art performance \cite{RN201,yang2017quicksilver,HU20181,gong2022uncertainty,hoopes2021hypermorph,hoffmann2020synthmorph,dalca2019learning,haskins2020deep,fu2020deep,de2019deep,mahapatra2018deformable,wang2020deepflash,deshpande2019bayesian,lei20204d}. For example, to accelerate the time-consuming conventional deformable registration process, Balakrishnan \textit{et al.} proposed the well-known VoxelMorph model to directly predict the so-called displacement vector field (DVF) for moving image warping \cite{RN232}. Despite the great computation efficiency of the \textit{general} VoxelMorph model that is trained from a general population dataset, its performance is usually sub-optimal for a specific testing case, especially when the testing case is out-of-distribution of the training dataset. To alleviate this issue, the test time optimization (TTO) concept \cite{9561808,chen2020generating,fechter2020one} was introduced, whose core idea was to fine tune the above pretrained general model such that it can adapt to the new testing case. 

In theory, by using TTO, one can always overfit any specific testing case to within any precision by using a large enough neural network if we consider it as function fitting task (using the neural network to fit the DVF). This is guaranteed by the so-called \textit{universal approximation theorem} of neural network \cite{hornik1989multilayer}. However, in practice, we find that although the current TTO method can refine the overall registration result to some extent at the initial optimization stage, its performance will saturate afterwards despite there still exist some clinically-relevant regions whose registration results are not acceptable, especially when the fixed/moving images are in low-quality, such as the commonly used phase-resolved CBCT images in radiotherapy. \textit{We hypothesize that this performance saturation phenomenon can be explained by that the total gradient is overwhelmed by those gradients coming from the mis-registrations of the potential artifacts and/or some clinically irrelevant tissues, and thereby the gradient associated with the clinically relevant region is relatively too small to be used to guide the correction for these clinically relevant regions.}

To mitigate the above challenge, we first propose to subdivide the TTO method into two different optimization levels: full image-specific optimization (ISO) and region-specific optimization (RSO). The ISO level is the same as the current well-known TTO method. As for the RSO level, we render it as an \textit{interactive registration} process: the clinician will review the current registration result and, if necessary, indicate the region that is not acceptable by providing the associated bounding box; then further specific optimization targeting at that region will be conducted by constraining the loss calculation region and thus enforcing the gradient is dominated by the region-of-interest (ROI). Based on the above two different TTO levels split, for both efficiency and efficacy consideration, we further render a new DL-based image registration method as a three-steps workflow, where the general model, the ISO process and the RSO process will be applied sequentially as necessary. We summarize our contributions as follows:
\begin{itemize}
    \item We identify the potential reasons explaining the performance saturation of the popularly used TTO-based algorithm for image-specific optimization.
    \item For the first time, we further subdivide the TTO algorithm into ISO (image) and RSO (region) levels. We then propose a RSO-based deep interactive registration algorithm targeting at the refinement of the ROI indicated by the clinician.
    \item We propose a new three-steps DL-based image registration workflow which can gradually refine the registration results from both the image- and region-levels.
    \item Experimental results show that the proposed RSO-based method outperforms the conventional ISO-based method qualitatively and quantitatively.
\end{itemize}

\section{Related Work}
\subsection{DL-based medical image registration}
The DL-based image registration models can usually be divided into two parts: 1) DVF prediction module which can be modeled by a convolutional neural network (CNN); and 2) spatial transformation module \cite{jaderberg2015spatial} which can be used to warp the moving image based on the predicted DVF. Balakrishnan \textit{et al.} proposed the well-known VoxelMorph method that can be trained in an unsupervised fashion \cite{RN232}, which was then combined with a probabilistic generative model for diffeomorphic registration \cite{dalca2019unsupervised}. Zhao \textit{et al.} built a recursive cascaded network to further improve the performance of the unsupervised DL methods \cite{zhao2019recursive}. Kuang \textit{et al.} presented another unsupervised registration algorithm, ``FAIM'', which employs a new network architecture and a Jacobian determinants-based penalty loss, reaching higher accuracy with less trainable parameters \cite{10.1007/978-3-030-32692-0_74}. Note that all these DL-based models are trained on a population dataset, and deployed/tested in the conventional single forward network propagation manner. Our work focuses on the testing phase, and thus can be combined with them.
 
\subsection{Test time optimization}
The TTO method is used to optimize a model during the testing stage rather than the training stage, such that the model can adapt to the specific testing case. In the work of Chen \textit{et al.} \cite{chen2020generating} and Fechter \textit{et al.} \cite{fechter2020one}, TTO is applied to untrained DL-based models to generate anthropomorphic phantoms and to track periodic motion, respectively. Another work applied TTO to pre-trained general models for CBCT-CT registration in adaptive radiotherapy \cite{liang2022segmentation}. These TTO methods can improve the performance of the general model, while they are applied in the full image level. By contrast, we subdivide the TTO method into the image-specific level and the region-specific level, and further propose a three-steps testing workflow that allows both image- and region-levels refinement, where the region can be interactively specified by the clinician.

\section{Methods}
\label{section:methods}
\subsection{Notations and Problem Formulation}
\label{subsection:problem_formulation}
Mathematically, given one fixed image $I_F(x)$ and one moving image $I_M(x)$, the image registration task is to find a transformation mapping $\mathcal{T}$ that makes the transformed moving image is anatomically spatially aligned with the fixed image, i.e., 
\begin{equation}
    I_M(\mathcal{T}(x)) = I_F(x).
\end{equation}

Specifically, regarding the DL-based methods, we can firstly employ a CNN $\phi_w$ parameterized by $w$ to predict the DVF by feeding into the network a two-channel input composing of the fixed/moving images. We can then use the spatial transformation layer (STL) $S$ to warp the moving image based on the above predicted DVF.

Giving a training dataset $\{(I_F^i(x), I_M^i(x)), i\in (0, 1, 2, \cdots, N - 1)\}$ that consists of $N$ samples indexed by $i$ representing a population of patients' data, the training process is to find the weights $w$ that can minimize a predefined dissimilarity-based cost function, i.e.,
\begin{equation}
    w = \arg \min_w\sum_{i=0}^{N-1} \mathcal{L}(I_F^i(x) - I_M^i(\mathcal{T}(x))), \mathrm{where~}\mathcal{T} = S \circ \phi_w^G(I_F^i(x), I_M^i(x)),
    \label{eq: cost function-general}
\end{equation}
where $\circ$ denotes the STL $S$ is applied on the predicted DVF for spatial sampling. Since this model is trained with a population of patients and thus can generally be applied on any patients within the data distribution, we name the resultant model as general model $\phi_w^G$. 

Given a pair of fixed/moving images associated with a specific testing patient, we can further fine-tune the above general model $\phi_w^G$ by optimizing the following cost function:
\begin{equation}
    w = \arg \min_w L(I_F(x) - I_M(\mathcal{T}(x))), \mathrm{where~}\mathcal{T} = S \circ \phi_w^I(I_F(x), I_M(x)).
    \label{eq: cost function-patient}
\end{equation}

This is the so-called TTO technique. Since cost function~(\ref{eq: cost function-patient}) is optimized in a full image specific level, we name the resultant model as ISO model $\phi_w^{I}$.

Comparing equations~(\ref{eq: cost function-general}) and~(\ref{eq: cost function-patient}), it is reasonable to conjecture that the reason that the ISO model $\phi_w^I$ outperforms the general model $\phi_w^G$ is that the general model has to balance the fitting degrees of all the training samples when minimizing the cost function~(\ref{eq: cost function-general}), while the ISO model just need to overfit the specific testing patient.

In theory, given a large enough CNN $\phi_w$, it can approximate the underlying real transformation map $\mathcal{T}$ to any small error $\epsilon$, i.e., $|\mathcal{T} - S \circ \phi_w| < \epsilon$. This is guaranteed by the well-known \textit{universal approximation theorem} of neural network \cite{hornik1989multilayer}. 

However, in practice, during the minimization of the cost function~(\ref{eq: cost function-patient}) with respect to the ISO model, we found that the loss decreases slowly after the initial stage's optimization (as would be demonstrated in Figures~(\ref{fig:result2}) and~(\ref{fig:result1})), despite some parts of the warped image are still far from perfect compared to the fixed image. This would reduce its clinical value when those parts correspond to some clinically relevant regions since the clinician cannot wait until it is fully converged, not to mention sometimes the converged result is still not acceptable.

We hypothesize that the above phenomenon is caused by the divergence of the gradient directions originated from different regions of the image. Mathematically, let's reformulate equation~(\ref{eq: cost function-patient}) as follows:
\begin{equation}
    w = \arg \min_w \sum_k^{K-1}\mathcal{L}(I_F(x^{r_k}) - I_M(\mathcal{T}(x^{r_k}))), \mathrm{where~}\mathcal{T} = S \circ \phi_w^I(I_F(x), I_M(x)),
    \label{eq: cost function-region}
\end{equation}
where $x^{r_k}$ represents the $r_k$th region of the image $x$. Without loss of generality, we assume that all the image regions satisfy the conditions that $\cup_{k=0}^{K-1}x^{r_k}=x$, $x^{r_i}\cap x^{r_j}=\emptyset,  \forall i, j \in {0,1,\cdots, K-1}, i \neq j$, $\cup$ and $\cap$ denote the union and intersection of two sets, respectively.

If one chooses the gradient descent-based optimizer to minimize the cost function defined in equation~(\ref{eq: cost function-region}), the weights $w$ can be updated 
as follows:
\begin{equation}
w = w - \lambda \frac{\partial \mathcal{L}}{\partial w} = w - \lambda \sum_{k=0}^{K-1} \frac{\partial \mathcal{L}^k}{\partial w}
\end{equation}
where $\frac{\partial \mathcal{L}^k}{\partial w}=\frac{\partial \mathcal{L}(I_F(x^{r_k})- I_M(\mathcal{T}(x^{r_k})))}{\partial w}$ represents the gradient contributed by the image region $x^{r_k}$. It is easily to derive that if the registration result associated with the region $x^{r_k}$ is not acceptable yet, while its associated gradient $\frac{\partial \mathcal{L}^k}{\partial w}$ cannot dominate the overall gradient $\frac{\partial L}{\partial w}$, representing the updating direction of the weights, or is even overwhelmed by the gradients from the other image regions, the registered result associated with this region will be hard to be refined even given more iterations. The above scenario happens frequently when there exist many artifacts in either fixed or moving images, which are hard, if not impossible, to be registered.

\subsection{Region specific optimization-based deep interactive registration}
\label{subsection:interactive_registration}
To mitigate the above challenge, in this work, we propose a region specific optimization (RSO)-based deep interactive registration algorithm that can target at the specific optimization of the ROI by constraining the loss and hence the gradient calculation region. The cost function of our proposed RSO method can be expressed as:
\begin{equation}
    w = \arg \min_w \mathcal{L}(I_F(x^{r_k}) - I_M(\mathcal{T}(x^{r_k}))), \mathrm{where~}\mathcal{T} = S \circ \phi_w^R(I_F(x), I_M(x)).
    \label{eq: cost function-RSO}
\end{equation}

We name the resultant model as region-specific model $\phi_w^{R}$. The workflow of our RSO method is as follows: a clinician reviews the current registration result; if the clinician finds that a clinically relevant ROI is not acceptable, the clinician will indicate the location of the ROI by providing the associated bounding box, and then our RSO method will be employed to refine the above ROI by optimizing cost function~(\ref{eq: cost function-RSO}). The above process can repeat until an acceptable registration result is achieved. 

\subsection{Three-steps DL-based image registration workflow}

\begin{figure}[h]
    \centering
    \includegraphics[width=1\textwidth]{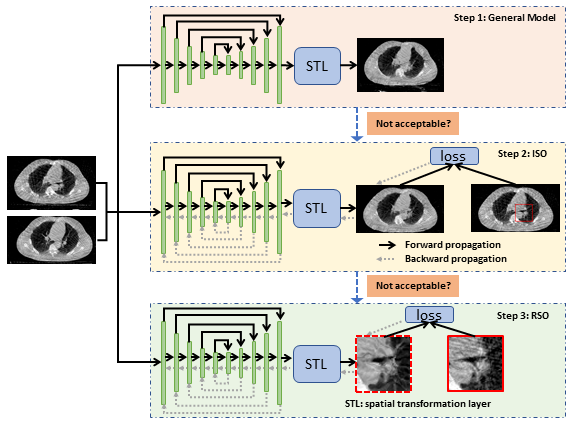}
    \caption{Three-steps DL-based image registration workflow.}
    \label{fig:method}
\end{figure}

Given their own advantages of the general model $\phi_w^G$ (fast inference), the ISO process (quick image level refinement) and the RSO process (accurate and fast region level refinement), for both efficiency and efficacy considerations, as demonstrated in Figure~(\ref{fig:method}), we further envision a new DL-based image registration method into a three-steps workflow, and detail it in Algorithm~(\ref{alg: algorithm}):
\begin{enumerate}
    \item The general model is used to generate an initial registration result;
    \item The clinician will then review the above result, and the ISO will be applied for a rough image level refinement if necessary;
    \item The clinician will further review the results specifically focusing on the clinically relevant regions, and if necessary, the RSO will be applied for a fine region level refinement by indicating the location of the region.
\end{enumerate}

\begin{algorithm}
\label{alg: algorithm}
\caption{Pipeline of the proposed three-steps deep learning-based medical image registration. (Our code will be open sourced once this work is accepted for publication.)}
\KwData{Fixed/Moving Images}
\KwResult{DVF and Warped Image}

Initialization: Pretrained general model $\phi_w^{G}$, Adam optimizer with constant learning rate $3\times 10^{-4}$ \;
The clinician checks the result from step 1 \;
\uIf{The result is good enough}{return\;}
\uElseIf{The general image quality is not acceptable}{Go to step 9 for ISO\;}
\uElseIf{The general image quality is acceptable while part of the regions is not acceptable}{Go to step 14 for RSO\;}
Image specific optimization (ISO) by optimizing equation (3) for 100 iterations  \;
\uIf{The result is good enough}{return \;}
\uElseIf{Some regions are not acceptable}{Go to step 14 for RSO\;}
Region specific optimization (RSO) based on the clinician provided bounding box by optimizing equation (6) for 400 iterations \;
Repeat Step 14 until all the regions are acceptable. 
\end{algorithm}

\section{Experiments and Results}
\subsection{Data and Implementation}
\label{subsection:dataset}
In this work, we use the open-accessed 4D Lung CT dataset \cite{hugo2017longitudinal} to validate the proposed method. Specifically, this dataset consists of the 4D fan-beam CT (FBCT) images and cone-beam CT (CBCT) images that were acquired based on 20 locally advanced, non small cell lung cancer patients. Each patient may contain multiple FBCT/CBCT images collected from different dates. In this work, without loss of generality, we employ the CBCT patient data for the performance validation. These 20 patients were then randomly split into 16 and 4 patients, serving as the training and testing datasets.

Since our method can be combined with any model architecture, without loss of generality, in this work, we adopt the VoxelMorph type of architecture as our general model, whose input has two channels (fixed/moving images) with a spatial dimension of $128\times 256 \times 256$.

To our knowledge, this is the first work on DL-based interactive registration, and thereby there is no existing dataset can be used to measure the performance. Considering this fact, in this work, we visually identify two regions from two different patient CBCT images that show obvious mis-registration errors even after many ISO-based refinement iterations. We will compare the registration results after general registration, 500 ISO-based refinement iterations, and our proposed method (100 ISO-based refinement iterations plus 400 RSO-based refinement iterations). 

We will visually compare the registration results and also use root mean squared errors (RMSE) to quantitatively compare the associated reconstruction errors between the fixed image and the warped images.

\label{section:Appendix}
\begin{table}[h]
    \centering
    \begin{tabular}{m{15em}m{15em}}
        \hline \hline
        Input Channels & 2 \\
        Input Size & $128\times 256 \times 256$ \\
        Output Channels & 3 \\
        Output Size & $128\times 256 \times 256$ \\
        Feature Numbers & 16, 32, 64, 128, 256 \\
        Strides & 2, 2, 2, 2 \\
        Residual Units/Layer & 2 \\
        \hline \hline
    \end{tabular}
    \caption{Without loss of generality, we used the VoxelMorph type of architecture as the testing framework, which is consisting of two major modules, i.e., the DVF estimation module and the spatial transformation module. Specifically, we used the MONAI implemented 3D U-Net architecture (\url{https://docs.monai.io/en/stable/_modules/monai/networks/nets/unet.html}) to predict the DVF. Regarding the spatial transformation module, we used the implementation from the open-sourced VoxelMorph code (\url{https://github.com/voxelmorph/voxelmorph/blob/dev/voxelmorph/torch/layers.py}).}
    \label{tab: network}
\end{table}

\begin{table}[h]
    \centering
    \begin{tabular}{m{20em}m{15em}}
    \hline \hline
       Normalization  & $(x + 1000) / 2000 \mathrm{~for~each~voxel~} x$ \\
       Rotation Probability & 100$\%$ \\
       Rotation magnitude along x, y, z-axis (rad) & $\pm0.01$, $\pm0.01$, $\pm0.05$ \\
       Zoom Probability & 100\% \\
       Zoom Magnitude & $0.9\sim 1.1$ \\
       \hline \hline
    \end{tabular}
    \caption{During the training of the general model, we first randomly choose one patient from the training dataset, we then randomly select two CBCT images from two different phases which might come from one same or two different collection dates, serving as the fixed/moving images. Random data augmentations with parameters listed in the table were applied to enlarge the richness of the dataset. We finally apply the above normalization strategy to preprocess the input images before feeding them into the network. During the testing phase, including the ISO/RSO processes, we only apply the above normalization strategy for preprocessing.}
    \label{tabl: data}
\end{table}

\begin{table}[]
    \centering
    \begin{tabular}{m{10em}m{30em}}
    \hline \hline
      Batch size & 1 \\
      Initial learning rate & $3\times 10^{-4}$ \\
      Total iterations & $1\times 10^5$ \\
      Learning rate policy & reduced to $3\times 10^{-5}$/$3\times 10^{-6}$ at $5\times 10^{4}$/$7.5\times 10^{4}$ iterations, respectively \\
    \hline \hline
    \end{tabular}
    \caption{Training parameter settings for the general model training. }
    \label{tab: training}
\end{table}

The readers are referred to Tables~(\ref{tab: network}),~(\ref{tabl: data}) and ~(\ref{tab: training}) for more experimental setting details.

\subsection{Results}

\begin{figure}[h]
    \centering
    \includegraphics[width=1\textwidth]{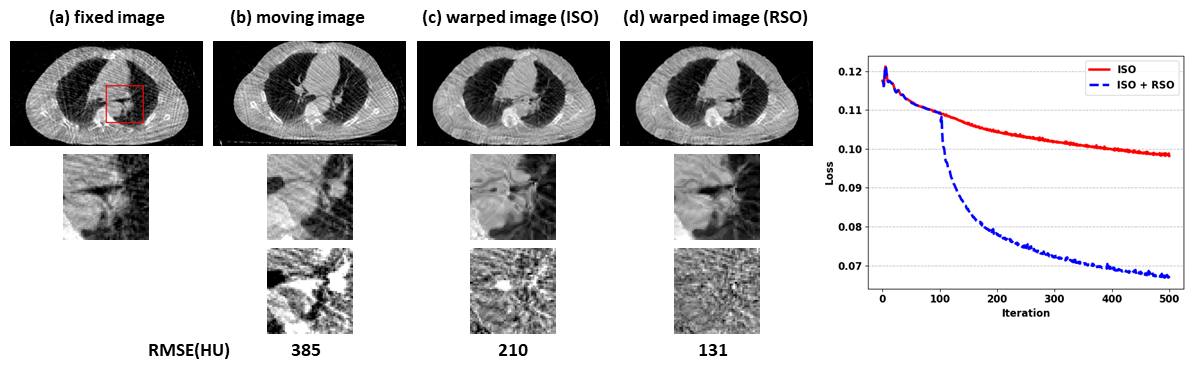}
    \caption{Registration results associated with the investigated case 1. (a) fixed image, (b) moving image, (c) and (d) warped images with respect to the image specific optimization (ISO) method, and our proposed method that combining ISO and region specific optimization (RSO). The second row shows the zoomed-in views of the red box. The third row shows the associated difference images compared with the fixed image. The root mean squared error (RMSE) is listed in the bottom with the unit of HU. Display windows: [-1000, 500]HU for the first two rows, and [-500, 500]HU for the third row. The loss dynamics for both methods are depicted in the right.}
    \label{fig:result2}
\end{figure}

\begin{figure}
    \centering
    \includegraphics[width=1\textwidth]{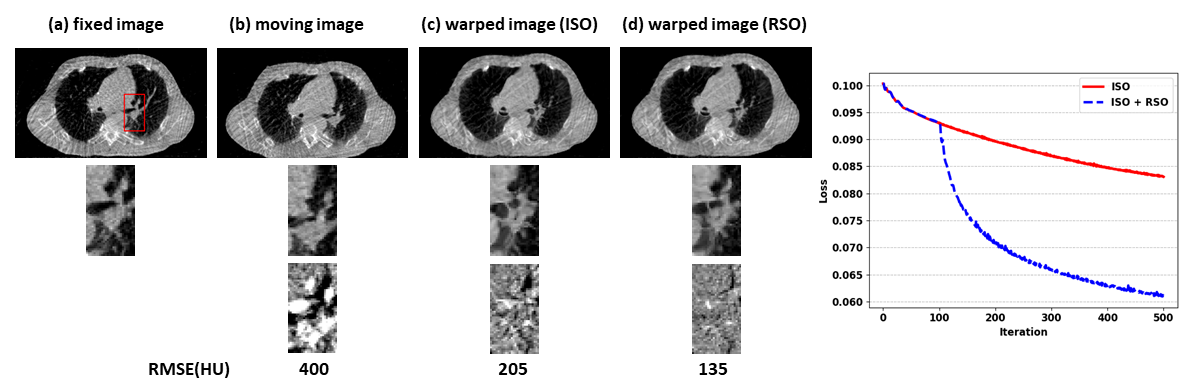}
    \caption{Registration results associated with the investigated case 2. Same sub-figures arrangement and display windows as Figure~(\ref{fig:result2}) are used in this case.}
    \label{fig:result1}
\end{figure}

We show the results associated with the two investigated cases in Figures~(\ref{fig:result2}) and (\ref{fig:result1}), respectively. Strong streaking artifacts due to the phase-resolved projection undersampling  can be observed in both the fixed/moving images. These streaking artifacts substantially increase the registration difficulty since there are no anatomically spatial correspondence for them among the CBCT images from different phases. Thus, during the ISO process, the gradient might be dominated or even overwhelmed by those streaking artifacts,  leading to some regions with sub-optimal registration results, as indicated by the red boxes in the fixed images. This phenomenon can be clearly observed  by comparing the zoom-in regions that associated with the fixed images and the ISO warped images, as shown in the second rows of Figures~(\ref{fig:result2}) and (\ref{fig:result1}). By excluding those clinically non-relevant regions during the gradient calculation, our proposed RSO method can substantially improve the registration results, as demonstrated by the zoomed-in ROIs. To have a better comparison, we also demonstrate the difference images between the fixed images and the other three images. We can find that the difference images associated with our proposed RSO method are dominated by the noise, while there remain strong structures in the difference images with respect to the ISO warped images. Quantitatively speaking, the ISO method can reduce the averaged CT value from 385/400HU to 210/205HU, while our RSO method can further reduce the error to 131/135HU, corresponding to the cases in Figures~(\ref{fig:result2})/(\ref{fig:result1}), respectively. In addition, from both loss dynamics, we can easily find that the loss associated with the ISO method (red solid line) decreases slowly, while our proposed RSO method (blue dash line), starting from iteration 100, decreases much faster. 

\section{Discussions and Conclusions}
This work is part of our ongoing effort towards developing AI- and clinician-integrated systems (AICIS) in medicine. In this study, we designed a new deep interactive registration method such that the clinician can work with the model
collaboratively to achieve a clinically acceptable result in an efficient manner.

To ensure a smooth user experience, fast registration result updating is desirable. In this work, to prove the concept of our RSO-based deep interactive registration method, we simply used the whole fixed/moving images as the input. However, in practice, to reduce the computation burden, we can only feed into the model a much smaller ROI, whose rough coordinates can be easily derived from the existing DVF of that ROI.  

In this work, because there is no existing dataset that can be used readily for our model's performance evaluation, we only used two showcases to demonstrate our algorithm, where the two ROIs are selected manually. This is one of the major limitations of our current work. In the future, we will work with the clinicians to identify and create a large enough dataset for a more comprehensive evaluation and analysis of our method. 

In summary, we demonstrate the concept of a three-steps DL-based interactive image registration method, where the clinicians review and indicate the region that is not acceptable, and then our proposed RSO method will be employed for specific optimization. Experimental results showed that our method outperforms the popular ISO method by reconstructing more faithful anatomical structures.

\bibliographystyle{splncs04}
\bibliography{./ref}

\end{document}